\documentclass[useAMS,usenatbib,fleqn]{nature}
\usepackage{graphicx}
\makeatletter
\let\saved@includegraphics\includegraphics
\AtBeginDocument{\let\includegraphics\saved@includegraphics}
\renewenvironment*{figure}{\@float{figure}}{\end@float}
\usepackage{calrsfs}
\DeclareMathAlphabet{\pazocal}{OMS}{zplm}{m}{n}
\usepackage{multirow}
\usepackage{hhline}
\usepackage{times}
\usepackage{graphics,epsf}
\usepackage[T1]{fontenc}
\usepackage{aecompl} 
\usepackage{amsmath}               
\usepackage{placeins}
\usepackage{booktabs}
\usepackage{amsfonts}               
\usepackage{amssymb}                
\usepackage{epsfig}                 
\usepackage{epstopdf}
\usepackage{rotating}
\usepackage{color}
\usepackage{tabularx}
\usepackage{url}

\def \apj{ApJ}
\def \aap{A\&A}
\def \aj{AJ}

\def \mnras{MNRAS}
\def \apjl{ApJ Lett.}

\def \nat{Nature}

\def \araa{ARA\&A}

\def \araa{ARA\&A}

\def \icarus{Icarus}

\def \planss {Planetary and Space Science}

\begin{document}



\title{Origin of (2014) $\rm MU_{69}$-like Kuiper-belt contact binaries from wide binaries}

\author{Evgeni Grishin$^{1}$, Uri Malamud$^{1}$, Hagai B. Perets$^{1}$, 
Oliver Wandel$^{2}$, Christoph M. Sch{\"a}fer$^{2}$}

\maketitle

\begin{affiliations} 

\item Physics Department, Technion - Israel Institute of Technology, Haifa,
Israel 3200002

\item Institut f{\"u}r Astronomie und Astrophysik, Eberhard Karls Universit{\"a}t T{\"u}bingen, Auf der Morgenstelle 10, 72076 T{\"u}bingen, Germany

\end{affiliations} 

\maketitle
\begin{abstract}
 Following its flyby and first imaging the Pluto-Charon binary, the New Horizons spacecraft visited the Kuiper-Belt-Object (KBO) (486958) 2014 MU$_{69}$ (Arrokoth). Imaging showed MU$_{69}$ to be a contact-binary, made of two individual lobes connected by a narrow neck, rotating at low spin period ($15.92\ \rm h$), and having high obliquity ($\sim98^{\circ}$)\cite{SternEtAl-2019}, similar to other KBO contact-binaries inferred through photometric observations\cite{LAcerda2011}. The origin of such peculiar configurations is puzzling, and all scenarios suggested for the origins of contact-binaries\cite{gls02,rw2006,pn09} fail to reproduce such properties and their likely high frequency. Here we show that semi-secular perturbations\cite{ap12,gpf18} operating only on ultra-wide ($\sim0.1-0.4$ Hill-radius\cite{grish17}) KBO-binaries can robustly lead to gentle, slow-speed binary mergers at arbitrarily high obliquities, but low rotational velocities, that can reproduce MU$_{69}$'s (and similar oblique contact binaries) characteristics. Using N-body simulations, we find that $\sim 15\%$ of all ultra-wide binaries with cosine-uniform inclination distribution\cite{pn09,npr10} are likely to merge through this process. Moreover, we find that such mergers are sufficiently gentle as to only slightly deform the KBO shape, and can produce the measured rotation speed of MU$_{69}$. The semi-secular contact-binary formation channel not only explains the observed properties of MU$_{69}$, but could also apply for other Kuiper/asteroid belt binaries, and for Solar/extra-solar moon systems. 
\end{abstract}

The discovery of MU$_{69}$'s bilobate shape and peculiar configuration opens new avenues of exploration and provide new clues on the physical processes that sculpture the Solar-System. Here we describe a novel evolutionary channel for the formation of MU$_{69}$ from an initially wide binary, producing MU$_{69}$-like objects. We consider the initial binary to be a member of a hierarchical triple together with the Sun. Due to secular evolution induced by the Sun, the inner orbit may experience changes in its eccentricity and mutual inclination on secular time-scales much longer than their orbital periods, known as  Lidov-Kozai (LK) oscillations, which can be modelled using a secular orbit averaging approach\cite{lid62,koz62}. Large LK-oscillations take place when the mutual inclination is large ($40^\circ \lesssim i\lesssim 140^{\circ}$). The highest eccentricities are attained as the binary evolves to the lowest inclinations and vice-versa\cite{naoz2016}.
 
If the binary eccentricity exceeds a threshold $e_{\rm coll}$, the small pericentre allows binary collisions. Thus, LK evolution could lead to coalescence of individual Kuiper-Belt-Binary (KBB) members into a {\rm single}, likely irregularly shaped KBO\cite{pn09}. However, since the closest approach occurs concurrently with the lowest inclinations, collisions mostly occur near $i \approx 40^\circ, 140^{\circ}$\cite{ft07}. Moreover, tidal effects and the non-spherical structure of the KBB components quench LK evolution, which makes collision possible only at a small part of the parameter space\cite{pn09,por+12}. The standard LK mechanism is therefore disfavoured for the origin of the highly-oblique MU$_{69}$, but can explain the origin of highly eccentric KBBs such as WW31 and  2001 QW322\cite{Vei+02, Pet+08, pn09}.

\begin{figure}
\begin{centering}
\includegraphics[width=8.5cm]{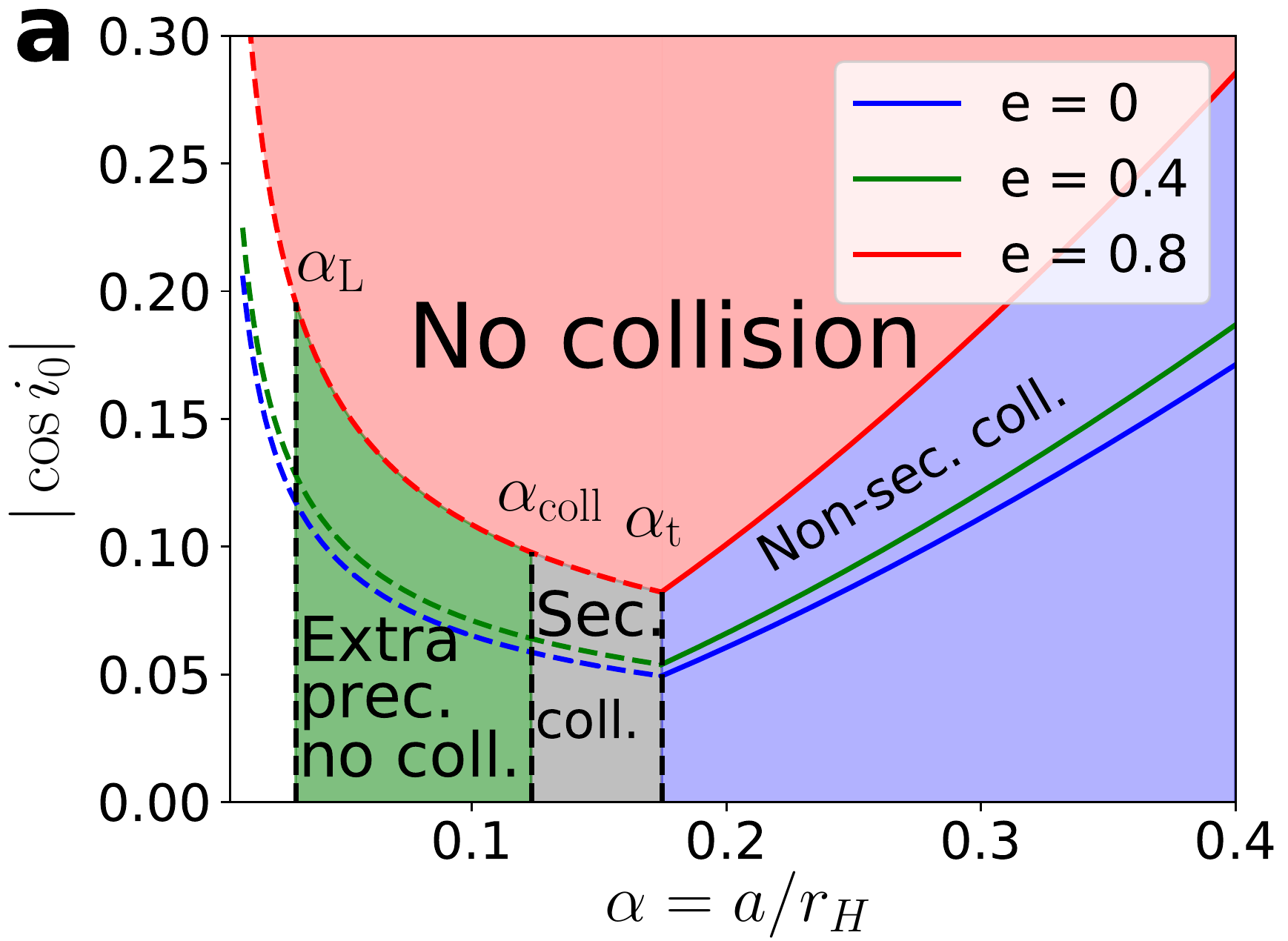}\includegraphics[width=8.3cm]{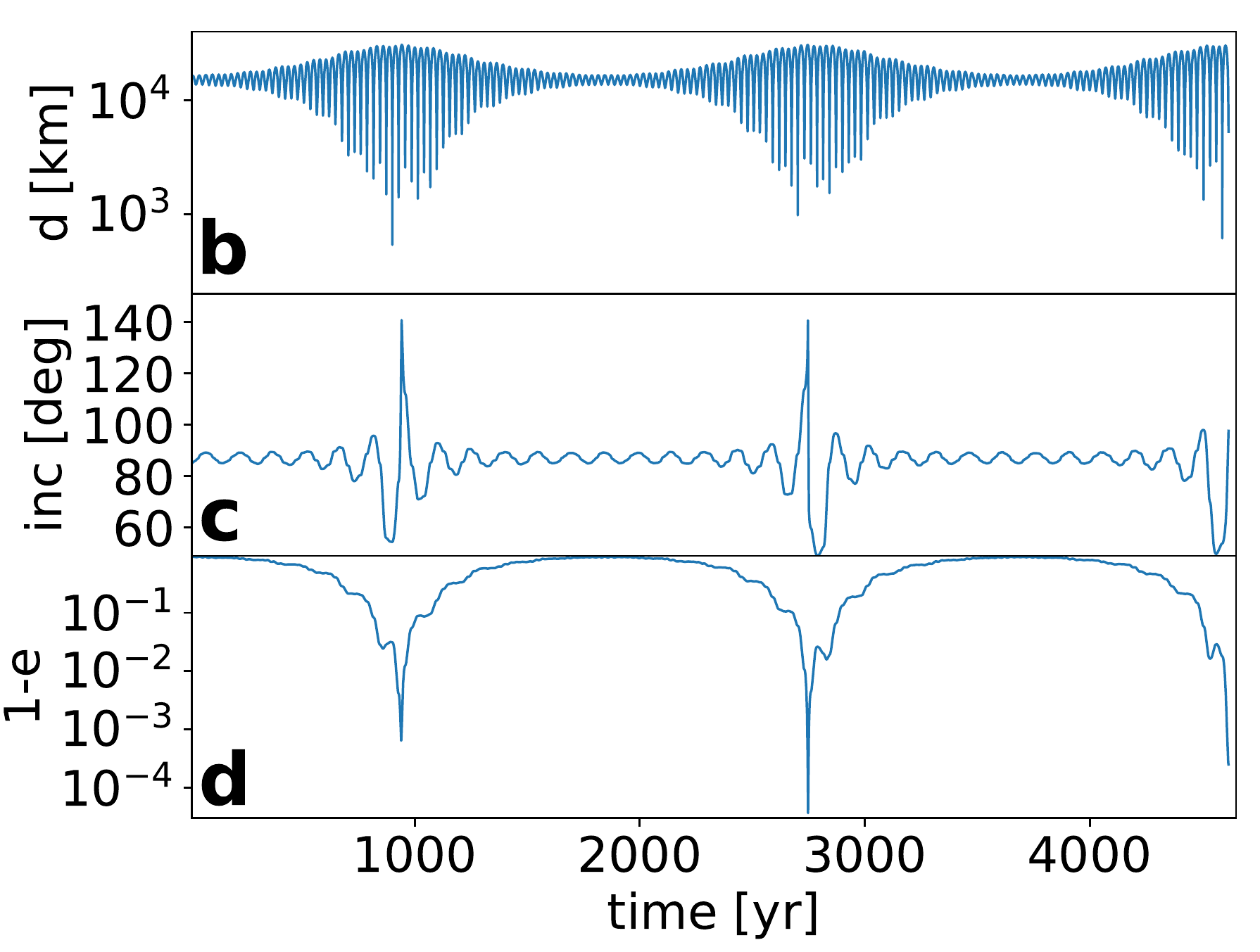}
\par\end{centering}
\caption{\label{fig:1}\textbf{| Roadmap to collisions of MU$_{69}$. (a),} The orbital evolution of MU$_{69}$ in initial separation/initial inclination space. The initial eccentricity is (from bottom to top lines) $0$ (blue), $0.4$ (green) and $0.8$ (red). Solid lines show the condition for a non-secular collisions (Eq.~\ref{eq:non_sec_cond}) with an unbound eccentricy. Dashed lines show the conditions for a secular collision (Eq.~\ref{eq:icrit}) and a deterministic eccentricity. The different domains are as follows: White - LK oscillations are completely quenched and the eccentricity is constant, $\alpha<\alpha_{{\rm L}}$ (Eq.~\ref{eq:alpha_L}). Green - the eccentricity is excited, but below $e_{\rm coll}$, and collisions are avoided, $\alpha_{{\rm L}}<\alpha<\alpha_{{\rm coll}}$ (Eq.~\ref{eq:alpha_coll}). Grey - secular evolution can lead to a collision, $\alpha_{{\rm coll}}<\alpha<\alpha_{{\rm t}}$ (Eq.~\ref{eq:alpha_t}). Blue - non-secular perturbations dominate and lead to a collision. Red - the initial inclination is too low to induce a collision. \textbf{(b)-(d)}, Time evolution of the instantaneous distance, inclination and eccentricity of an individual orbit with initial Keplerian elements of semi-major axis $a = 0.3 r_H$, eccentricity $e=0.1$, inclination  $i=86^{\circ}$, argument of periapse $\omega=0$, argument of ascending node $\Omega=\pi/4$ and mean anomaly $\mathcal{M}=0$. The outer binary is set at $\omega_{\rm out}=\Omega_{\rm out}=0$ and $\mathcal{M}_{\rm out}=-\pi/4$.}
\end{figure}

When the inner-to-outer period ratio increases, secular averaging breaks down, and the evolution is no longer secular, but only semi-secular. The inner-binary now evolves significantly during the outer-orbit timescale, and short-term fluctuations arise, making the LK evolution more complex\cite{ap12, luo16,gpf18}. The maximal eccentricity can be calculated analytically, including domains where it is unconstrained\cite{gpf18} and the evolution is non-secular. Fig. \ref{fig:1}\textbf{a} shows the analytical 2D parameter space for allowed and forbidden domains for collisions in terms of the initial inclination $\cos i_0$ and the initial separation of the inner binary normalized to the Hill radius $r_H = a_{\rm out} \left( m_{\rm in}/3M_{\odot} \right)^{1/3}$, where $\alpha\equiv a/r_H\lesssim\alpha_H=0.4$ is the Hill-stability limit for highly inclined orbits\cite{grish17}. We use the outer orbit parameters of MU$_{69}$: semi-major axis $a_{\rm out}=44.581\ \rm au$ and eccentricity $e_{\rm out}=0.041$. We model the lobes as triaxial ellipsoids of dimensions approximately $22 \times 20 \times7$ and $14 \times 14\times 10\ \rm km^3$ \cite{SternEtAl-2019}, leading to a total radius of $R_{\rm tot} = 18\ \rm km$ and inner mass of $m_{\rm in} = (1.61 + 1.03) \cdot 10^{18}\ \rm g = 2.64\cdot 10^{18}\ \rm g$ for a density of $\rho=1\ \rm g\ cm^{-3}$ (see others densities in Methods). Collisions occur only for sufficiently large critical inclination and wide initial separation, which overcomes LK quenching ($\alpha>\alpha_{\rm coll}$, see Methods). The transition from secular to non-secular dominated regimes is given by
\begin{equation}
\alpha_{\rm t}=3^{1/3}\left[\frac{128}{135}\frac{(1-e_{{\rm out}}^{2})}{\left(1+\frac{2\sqrt{2}}{3}e_{{\rm out}}\right)^{2}}\left(\frac{M_{{\odot}}}{m_{{\rm in}}}\right)^{1/3}\frac{R_{{\rm tot}}}{a_{{\rm out}}} \right]^{1/4}, \label{eq:alpha_t}
\end{equation}
or $\alpha_{\rm t}\approx0.174$ in our case. Fig. \ref{fig:1}\textbf{b} demonstrates the separation in the non-secular regime prior to the collision. During the high-e phase, there are about $\sim 10$ cycles where the distance drops below $10^3\ \rm km$. A collision occurs during the third LK cycle after about $\sim 4600\ \rm yr$. The mutual inclination flips its orientation during the high-e peak of the LK cycle (\textbf{c}).The eccentricity is essentially unbound and a collision eventually occurs (\textbf{d}).

In order to explore in detail the overall evolution and statistics of KBBs in the chaotic, non-secular regime, we defer to detailed N-body simulations, which provide us with the probability for collisions and post-collision characteristics. We use the publicly available code \texttt{REBOUND}\cite{ReboundMain}, with the \texttt{IAS15}\cite{ReboundIAS15} integrator (see Methods for details and stopping conditions). We integrate four sets of initial conditions in the non-secular regime. Three sets include initial separations of $\alpha=0.2, 0.3, 0.4$, and a fourth one with uniformly sampled separations. The orbital angles are sampled uniformly. The mutual inclination of observed binaries is cosine-uniform\cite{npr10}, therefore we follow cosine-uniform sampling with a cut-off at $|\cos i| \le 0.4$ (since lower inclinations cannot lead to a collision). For each case, we run 250 simulations (except for $\alpha=0.2$, for which we run 200 simulations and $|\cos i| \le 0.3$), each up to $5\times 10^4\ \rm yr$.

\begin{figure}
\begin{centering}
\includegraphics[width=15cm]{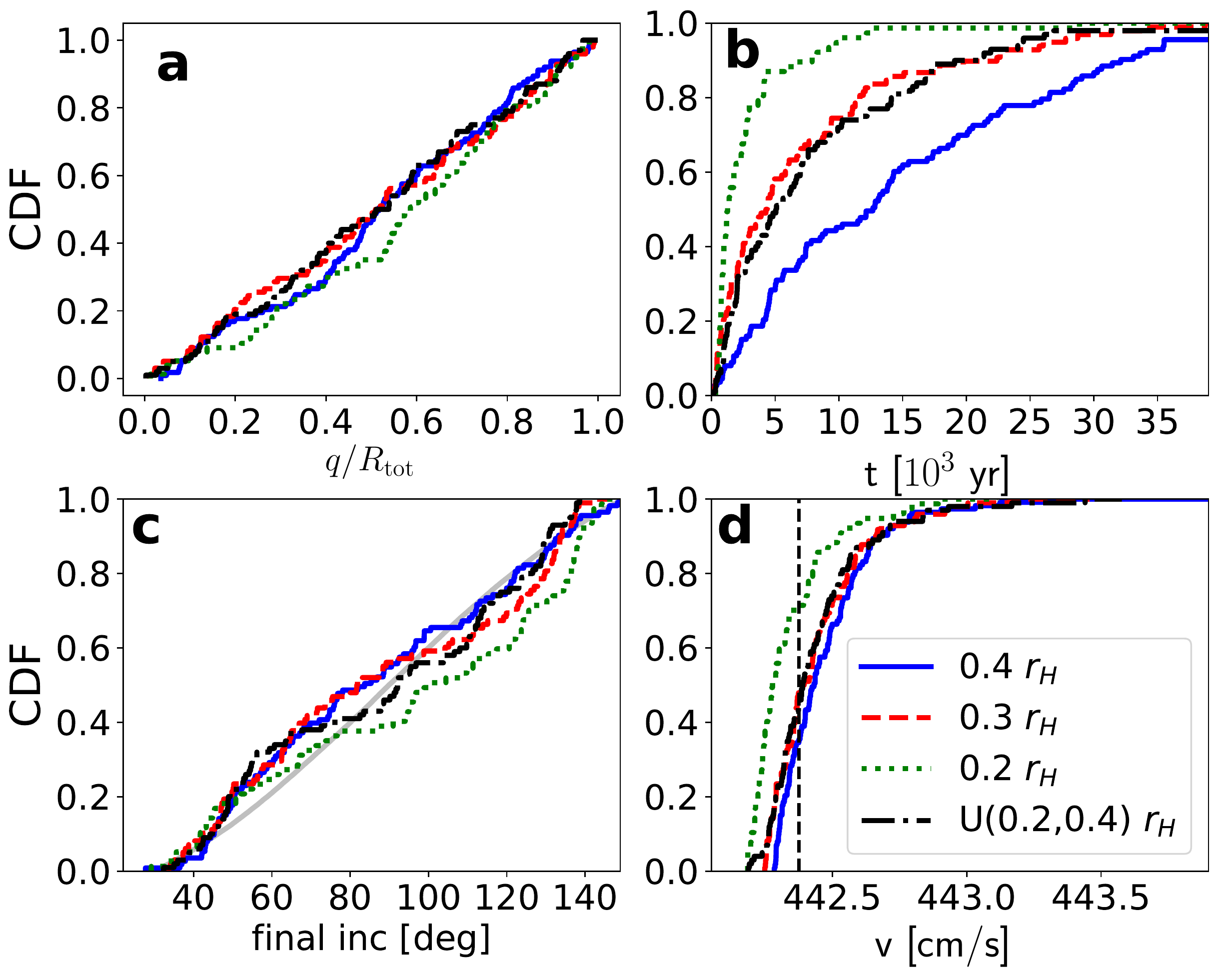}
\par\end{centering}

\caption{\label{fig:4}
 \textbf{| Cumulative distriubtions of the impact characteristics.} The CDFs are for $\alpha= 0.4$ (solid blue), $0.3$ (dashed red), $0.2$ (dotted green) and uniform in $(0.2, 0.4)$ (dash-dotted black).\textbf{(a),} Pericentre $q/R_{\rm tot}$. \textbf{(b),} Time of collision. \textbf{(c),} Final inclination at impact. The shaded grey line is a uniform cumulative distribution in $\cos i$ in the range of $30-150$ deg. \textbf{(d),} Velocity at impact. The vertical black dashed line is the escape velocity. }
\end{figure}

Fig.~\ref{fig:4} shows the cumulative distribution function (CDF) of various parameters of the colliding
orbits. Both the closest approach $q=a(1-e)/R_{\rm tot}$ (\textbf{a}) and the final inclinations (\textbf{c}) at collision are consistent with a uniform distribution (in $\cos i$) between $40^\circ$ and $140^\circ$,
suggesting that the orbits are indeed chaotic and in the non-secular regime, as expected. Most orbits induce collisions after about a few thousand years (\textbf{b}). The mean collision time increases with increasing separation. The velocity at impact is comparable to the escape velocity with a very small dispersion, consistent with a gentel collision\cite{Mckinnon2020} (\textbf{d}).

We find the overall merger fractions of wide-binaries to be around $12\% - 18 \%$ (see Extended Data Table 1), which is roughly consistent with the observed $10\% - 25\%$ contact-binaries for the cold classical belt\cite{TS2019}. Most mergers occur for initially high inclinations, as expected. A fraction of $1 \% - 3 \%$ of the overall wide binaries produce highly oblique contact-binaries ($80^{\circ}-100^{\circ}$), which are consistent with the observed high obliquity of $\rm MU_{69}$ and provide clear predictions for future KBO observations with larger statistics.
There is little dependence on the underlying distribution of $\alpha$, and the merger rates are bounded between the minimal and maximal values of $12\%$ for $\alpha=0.2$ and $18\%$ for $\alpha=0.4$, respectively. Moreover, in a collisional environment\cite{Par+12} the binary orbits can be perturbed such that originally low-inclination orbits become highly-inclined and be subjected to semi-secular evolution to form contact-binaries; the quoted formation rates are therefore only lowers limits to the total fractions of contact-binaries formed through this process.

The non-merging systems will continue to evolve quasi-periodically. On longer timescales, three-body encounters are expected to shape the populations of KBBs\cite{gls02,per2011}. Exchange interactions can drive the binaries into equal masses\cite{fun2004}, while the 'softness' of the binaries can make them become even softer and evaporate (Heggie's law\cite{heggie75}). There is only a handful of KBBs beyond $a\gtrsim 0.05\ r_H$ with either prograde or retrograde orbits, without highly inclined ones (see Fig 1 of ref.\cite{GRUNDY2019}), while the widest known binary 2001 QW$_{322}$ at $a\approx 0.2\ r_H$ is expected to disrupt within a billion years\cite{Pet+08}.

\begin{figure}
\includegraphics[width=8.2cm]{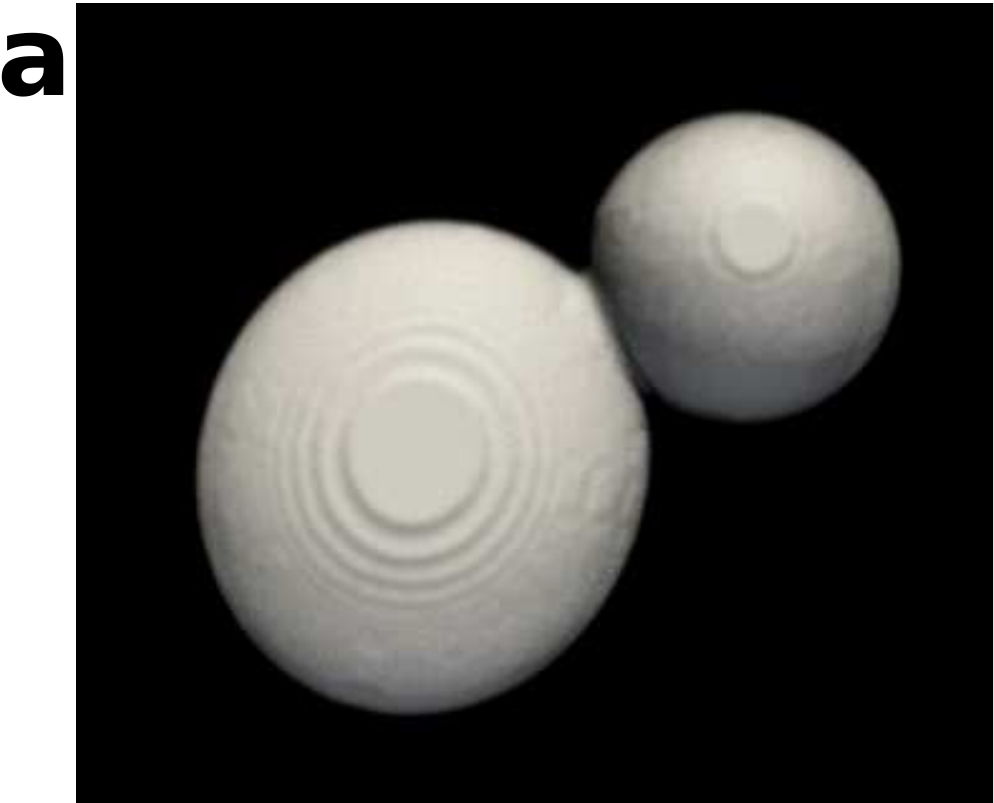}\includegraphics[width=8.5cm]{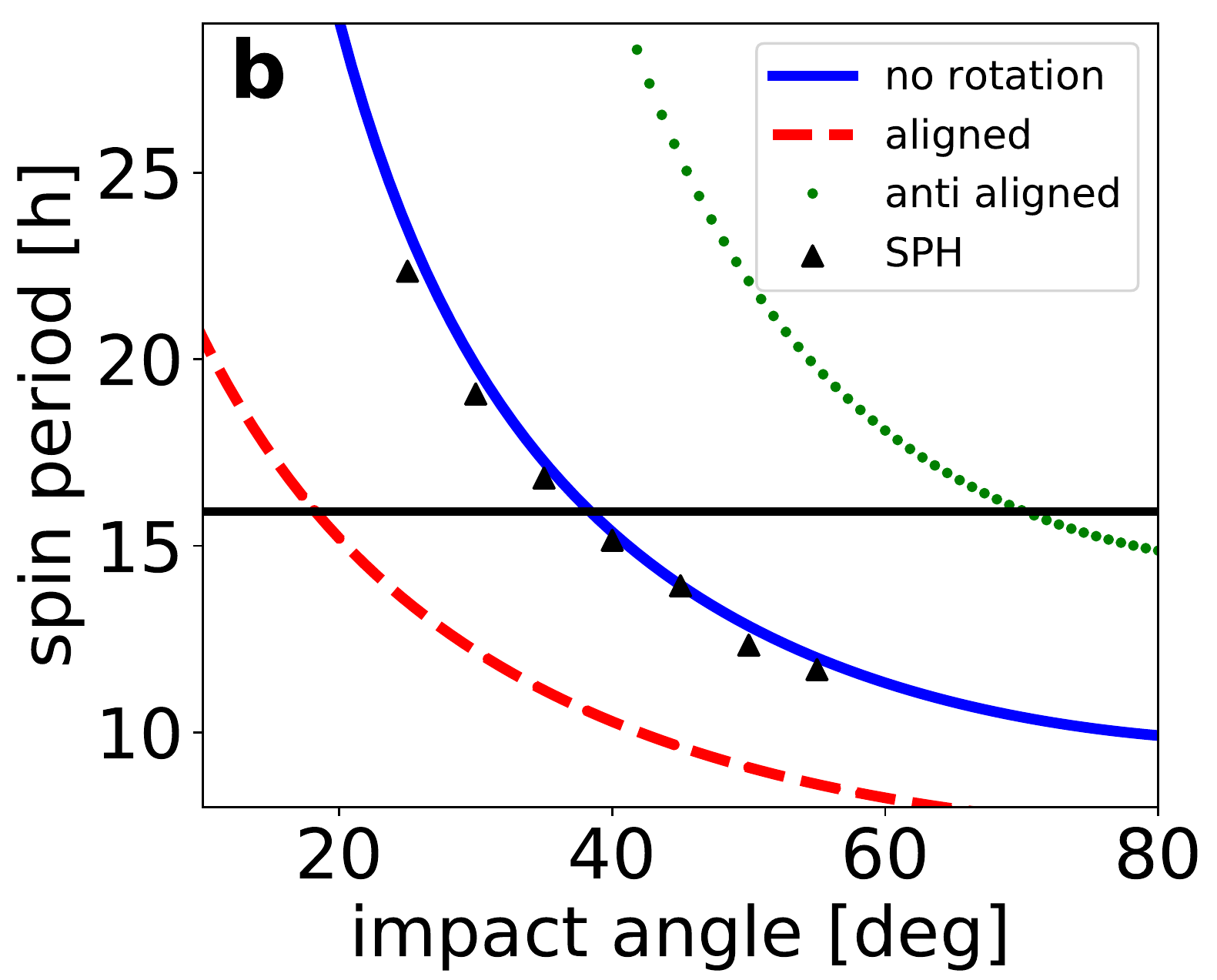}

\caption{\textbf{| Shape and spin period of MU$_{69}$. (a),} Final collision outcome at an impact angle of $40^{\circ}$ with high material strength (see Methods).  \textbf{(b),} Spin period as a function of the impact angle for MU$_{69}$. The horizontal line is the observed period of $15.92$ h. The solid blue line indicates initially non-rotating progenitors. The dashed red line indicates two initially aligned rotating progenitors with a period of $10\ $h, while the dotted green line indicates an anti-aligned configuration. The black crosses are the results obtained from SPH simulations (see Methods).\label{fig:spin}}
\end{figure}

In order to test the feasibility of MU$_{69}$'s semi-secular collision origin, we also need to account for the observed spin period of MU$_{69}$. Angular momentum conservation allows us to find the resulting spin period depending on the impact angle and the primordial spins of each component. The final impact parameter (which corresponds to an impact angle, see Methods) at collision is uniformly distributed, therefore our model can robustly produce a wide range of possible final rotations periods, without any fine-tuned modelling of MU$_{69}$ and its composition/density, and alleviate the angular momentum problem, unlike other models\cite{SternEtAl-2019}.

Fig. \ref{fig:spin}\textbf{a} shows the outcome of a collision at a $40^{\circ}$ impact angle with high material strength, which reproduces the shape of $\rm MU_{69}$. Low or medium-strength materials result in a deformed shape and are thus ruled out. If $\rm MU_{69}$'s density is halved compared to the fiducial $1 \rm g\  cm^{-3}$ value, as suggested by ref.\cite{SternEtAl-2019}, $v_{\rm esc}$ is lower, and in this case the medium-strength material also produces an undeformed shape. Random collisions,  even at relative velocities as low as $10v_{\rm esc}$ destroy or heavily deform the binary with high-strength materials, hence they are likewise ruled out (see Methods). Fig.\ref{fig:spin}\textbf{b} shows the expected spin period dependence on the impact angle. An impact angle of $\sim 40^{\circ}$ reproduces the observed spin period (see Methods) for initially non-spinning objects. Taking a typical initial spin period of $\sim 10\ \rm h$ with random orientations extends the range of plausible impact angles to $\sim 20^{\circ} - 70^{\circ}$, for maximally aligned and anti-aligned configurations, respectively. Smoothed Particle Hydrodynamics (SPH) collision simulations agree with our simplified estimate and support our assumptions of undeformed, rigid bodies when modelled with high-strength material parameters, or else medium-high-strength parameters if the density/impact-velocity is slightly lower (see ref.\cite{SchaferEtAl-2016} and Methods for details). 

Together, our dynamical and post-collisional modelling yields a consistent, coherent picture for the origin of MU$_{69}$ from an ultra-wide KBO-binary. Such wide KBB progenitors could be a natural by-product of KBO and KBB evolution in the early Solar-System\cite{gls02,gls04,N19}. Most likely the case of MU$_{69}$ is not unique, and secular/semi-secular evolution plays a major role in the evolution of many KBBs and the production of low-velocity collisions between individual KBB components. In fact, modelling of the impact origin of the Pluto-Charon system also suggests a low-velocity impact is required to explain its properties\cite{c2005}. Morevover, given the high obliquity of the Pluto-Charon system, it is possible that it also had originated from an initially wide-binary and followed a secular/semi-secular evolution similar to MU$_{69}$. Similar evolutionary scenarios might also apply to the evolution of other contact binaries such as (139775) 2001 QG298\cite{LAcerda2011} as well as moons and exo-moons, since all of them form hierarchical triple systems with their host star.



\textbf{Acknowledgements} We gratefully acknowledge useful discussions with Daniel C. Fabrycky and Edwin Kite. HBP acknowledges support from the MINERVA center for "Life under extreme planetary conditions" and the Kingsley fellowship at Caltech. CMS and OW acknowledge support by the High Performance and Cloud Computing Group at the Zentrum f{\"u}r Datenverarbeitung of the University of T{\"u}bingen, the state of Baden-W{\"u}rttemberg through bwHPC and the German Research Foundation (DFG) through grant no INST 37/935-1 FUGG. CMS acknowledges support from the DFG through grant no 398488521.

\textbf{Author information}
Correspondence and requests for materials should be
addressed to EG (eugeneg@campus.technion.ac.il) and HBP (hperets@ph.technion.ac.il). 
\section*{Methods}

\section*{Lidov-Kozai secular evolution }

Let us consider the evolution of a binary KBO due to LK secular evolution. Let the inner binary start with an initial separation of $a_{{\rm 0}}=\alpha r_{H},$
eccentricity $e_{0}$ and mutual inclination $i_{0}$. The Hill radius is $r_H = a_{\rm out} (m_{\rm in}/3M_\odot)^{1/3}$. The most stable
orbit is around $\alpha_{H}\approx0.4$.\cite{grish17} The minimal eccentricity
required for collision is
\begin{equation}
e_{{\rm coll}}=1-\frac{R_{{\rm tot}}}{\alpha r_{H}}. \label{eq:e_coll}
\end{equation}
Using the standard LK formula, the maximal eccentricity
is
\begin{equation}
e_{{\rm max}}^{2}=1-\frac{5}{3}(1-e_{0}^{2})\cos^{2}i_{0}. \label{eq:e_max}
\end{equation}
In order for a collision occur, we require that $e_{{\rm max}}\ge e_{{\rm coll}}$,
which yields the critical inclination
\begin{equation} \label{eq:icrit}
\cos i_{0}^{c} \approx\sqrt{\frac{6}{5(1-e_{0}^{2})}\frac{R_{{\rm tot}}}{\alpha r_{H}}} =\frac{0.037}{\sqrt{1-e_{0}^{2}}}\left(\frac{\alpha}{0.3}\right)^{-1/2},
\end{equation}
 where $R_{{\rm tot}}/r_{H}\ll1$ was expanded to linear order. The probability for collision can be expressed in terms of integrating over the distribution function $f_{a,\theta}(a,\theta)$ and where $\theta\equiv\cos i$.
For uniform, independent distribution in $\cos i$, as inferred from KBO observations\cite{npr10}, the probability is
\begin{equation}
P(e_{0})=\intop_{a_{{\rm min}}}^{a_{{\rm max}}}f_{a}(a)\theta_{c}(a)da, \label{eq:p2}
\end{equation}
where $\theta_c(a)$ is given by Eq. \ref{eq:icrit}.

\paragraph*{Inclination angle at impact:}
From the conservation of $j_{z}\equiv\sqrt{1-e^{2}}\cos i$
the inclination at impact is
\begin{equation}
\cos i_{{\rm coll}}=\frac{\sqrt{1-e_{0}^{2}}\cos i_{0}}{\sqrt{1-e_{{\rm coll}}^{2}}}=\sqrt{1-e_{0}^{2}}\cos i_{0}\sqrt{\frac{\alpha r_{H}}{2R_{{\rm tot}}}}. \label{eq:cosicoll}
\end{equation}

In order to get an impact angle of $i_{{\rm coll}}$,
we invert Eq. \ref{eq:cosicoll}:
\begin{equation}
    i_0 = \arccos \left(\cos i_{\rm coll} \sqrt{\frac{2R_{\rm tot}}{\alpha r_H}} \right)
\end{equation}
In order to get the observed obliquity $i_{\rm coll} = 98^{\circ}$, the critical inclination is at least
$|\cos i_{0}|\le 0.00585$, ($89.33^{\circ} \le i_0  \le 90.66^{\circ}$) for $\alpha=0.1$, which is unlikely. Moreover, for $\alpha\lesssim0.1$, collisions are unlikely regardless of the initial inclination due to the effects or oblateness, as shown below.

\paragraph*{Effects of oblateness:}
Small KBOs might not have spherical shapes, in which case their gravitational potential is not spherical. Such a configuration induces extra precession on the orbit which can significantly affect the secular evolution. The leading
term is encapsulated in the $J_{2}$ coefficient of the gravitational
potential\cite{md99}. Planets are mostly spherical, and their deviation is small;  $J_{2}\approx10^{-3}$ for Earth and around $J_{2}=0.014$
for Jupiter, and is related to the flattening of the planets induced by their rotations. In the case of MU$_{69}$ components, the objects are highly non-spherical and $J_{2}$ could be large. Using the principal moments of inertia of an oblate spheroid, we have $J_2 = (1-(c/a)^2)/5$, which is around $J_2 \approx 0.18$ for the primary and $J_2 \approx 0.1$ for the secondary.

The additional precession may quench the LK oscillations if it is too strong. In order to quantify the effects of the
additional precession we can define a dimensionless quantity that
measures the ratio between the LK and the oblateness induced precessions\cite{lml15,glp18}
\begin{equation}
\epsilon_{{\rm rot}}=\frac{3}{2}J_{2}\frac{m_{{\rm in}}}{m_{{\rm out}}}\frac{a_{{\rm out}}^{3}(1-e_{{\rm out}}^{2})^{3/2}R_{1}^{2}}{(\alpha r_{H})^{5}}, \label{eq:epsrot}
\end{equation}

Setting $\epsilon_{{\rm rot}}=3/2$ leads to the definition of the
Laplace radius\cite{T9,glp18} in terms of the Hill radius,
$r_{L}\equiv\alpha_{L}r_{H}$ where
\begin{equation}
\alpha_{L}  =\left(J_{2}\frac{m_{{\rm in}}}{m_{{\rm out}}}\right)^{1/5}\frac{\left(a_{{\rm out}}^{3}R_{1}^{2}\right)^{1/5}}{r_{H}}(1-e_{{\rm out}}^{2})^{3/10}
 =0.03\left(\frac{J_{2}}{0.2}\right)^{1/5}\left(\frac{R_{1}}{11\ {\rm km}}\right)^{2/5}, \label{eq:alpha_L}
\end{equation}
leading to $\epsilon_{\rm rot} = 1.5 (\alpha/\alpha_L)^{-5}$.\cite{lml15} showed that the maximal eccentricity attained is given by the implicit
expression for  $\cos i_{0}=0$ (their Eq.~50):
\begin{equation}
\frac{\epsilon_{{\rm rot}}}{3}\left(\frac{1}{(1-e_{{\rm max}}^{2})^{3/2}}-1\right)=\frac{9}{8}e_{{\rm max}}^{2} \label{eq:emaxrot}
\end{equation}
Expanding in $\epsilon_{{\rm rot}}\ll1$, $e_{{\rm max}}^{2}\approx1$
\begin{equation}
e_{{\rm max}}\approx1-\frac{2}{9}\epsilon_{{\rm rot}}^{2/3}. \label{eq:expanded_emax}
\end{equation}
A collision can occur only if $e_{{\rm coll}}<e_{{\rm max}}$ or
\begin{equation}
\alpha_{{\rm coll}}>\left(\frac{3}{2}\right)^{2/7}\alpha_{L}^{10/7}\left(\frac{2}{9}\frac{r_{H}}{R_{{\rm tot}}}\right)^{3/7}\approx0.12. \label{eq:alpha_coll}
\end{equation}

Note that a similar analysis can be made for tidal distortions or relativistic corrections\cite{lml15}. In this case, they are much weaker than the rotational effects.

\section*{Non-secular Lidov-Kozai  evolution}
In the previous section we considered the evolution due to secular LK evolution. In the semi-secular (semi-LK) regime\cite{ap12,gpf18},  short term fluctuations can significantly change the evolution. In the following we discuss the overall effects of such short-term perturbations. 
The strength of the perturbations is encapsulated in the single averaging parameter\cite{luo16,gpf18}:
\begin{equation}
\epsilon_{{\rm SA}}\equiv\left(\frac{a_{1}}{b_{{\rm out}}}\right)^{3/2}\sqrt{\frac{m_{{\rm out}}}{m_{{\rm in}}}}=\frac{\alpha^{3/2}}{\sqrt{3}(1-e_{{\rm out}}^{2})^{3/2}}\approx0.1\left(\frac{\alpha}{0.3}\right)^{3/2}. \label{eq:esa}
\end{equation}

One important quantity is the (averaged) z-angular momentum $\bar{j}_{z}=\sqrt{1-e_{0}^{2}}\cos i_{0}$,
assuming that $i_{0}$ and $e_{0}$ have their mean value.

The eccentricity of the orbit becomes unbound once the fluctuation in $j_{z}$, $\Delta j_{z}$ is larger than its initial value, namely $\Delta j_{z}>\bar{j}_{z}$.
The fluctuation is estimated analytically in\cite{luo16,gpf18},  and can be used to show that the eccentricity is unbound if
\begin{equation}
\cos i_{0}\sqrt{1-e_{0}^{2}}\lesssim\frac{9}{8}\tilde{\epsilon}_{{\rm SA}}\approx0.118\left(\frac{\alpha}{0.3}\right)^{3/2}, \label{eq:non_sec_cond}
\end{equation}
where $\tilde{\epsilon}_{{\rm SA}}=\epsilon_{{\rm SA}}\left(1+2\sqrt{2}e_{{\rm out}}/3\right)\approx1.039\epsilon_{{\rm SA}}$
is defined for convenience.

The width of the non-secular semi-LK regime increases with $\alpha$,
while the width of the secular LK regime decreases with increasing $\alpha$.
Comparing Eq.~(\ref{eq:icrit}) and Eq.~(\ref{eq:non_sec_cond}) yields the transitional separation $\alpha_{\rm t}$ found in Eq. \ref{eq:alpha_t}.

\section*{Spin period}

There is little evidence for structural changes of the object since its formation, and the spin period is believed to be primordial\cite{SternEtAl-2019}. In our model, the collision is gentle and occurs at relatively low velocities ($v_{\rm esc} = 442.4\ \rm cm \ s^{-1}$ for our nominal density and $v_{\rm escc} = 3.12.8\ \rm cm \ s^{-1}$ for the lower density of $0.5\ \rm g\ cm^{-3}$ assumed in ref.\cite{SternEtAl-2019}), such that almost any impact parameter (or impact angle) is allowed. Therefore, in order to obtain the observed spin period, we can use the standard arguments of angular momentum conservation and derive the impact parameter (or impact angle) that yields the desired spin rate.

Consider triaxial ellipsoidal bodies with masses $m_i$ and axes $a_i \ge b_i \ge c_i$, with $i=1,2$. We assume that the major axes $a_i$ are parallel, similar to the observed object, and that the collision occurs also in parallel with the major axes. The largest moment of inertia is $I_{3}^{(i)}=m_{i}(a_{i}^{2}+b_{i}^{2})/5$.

After the collision  the distance between the centre of masses of the joint body and each centre of the ellipsoid is $r_{i}$. Then the principal moment of inertia of the joint body is
\begin{equation}
I_{3}^{{\rm tot}}  =I_{3}^{(1)}+I_{3}^{(2)}+m_{1}r_{1}^{2}+m_{2}r_{2}^{2}\nonumber  =\sum_{i=1}^{2}\frac{m_{i}}{5}\left(a_{i}^{2}+b_{i}^{2}+5r_{i}^{2}\right).
\end{equation}

Now, the ellipsoids collide with relative velocity $v_{\rm esc}$ and impact
parameter $b$. The orbital angular momentum is $L_{z}=\mu bv_{\rm esc}$,
where $\mu=m_{1}m_{2}/(m_{1}+m_{2})$ is the reduced mass. If the
two bodies are non-rotating, then the joint angular frequency is 
\begin{equation}
\Omega=\frac{L_{z}}{I_{3}^{{\rm tot}}}=\frac{5\mu bv_{0}}{m_{1}\left(a_{1}^{2}+b_{1}^{2}+5r_{1}^{2}\right)+m_{2}\left(a_{2}^{2}+b_{2}^{2}+5r_{2}^{2}\right)}\label{eq:p}.
\end{equation}
If the individual bodies are rotating around the $z$ axis with frequencies
$\Omega_{i}$, the additional angular momentum of each body is $I_{3}^{(i)}\Omega_{i}$
for $i=1,2$, thus Eq. (\ref{eq:p}) becomes 
\begin{equation}
\Omega=\frac{5\mu bv_{0}+m_{1}(a_{1}^{2}+b_{1}^{2})\Omega_{1}+m_{1}(a_{1}^{2}+b_{1}^{2})\Omega_{2}}{m_{1}\left(a_{1}^{2}+b_{1}^{2}+5r_{1}^{2}\right)+m_{2}\left(a_{2}^{2}+b_{2}^{2}+5r_{2}^{2}\right)}.
\label{eq:p22}
\end{equation}

For an impact angle $\theta$, the distance of the point
of contact to the centre of each ellipsoid is $\xi_{i}=\text{\ensuremath{\sqrt{a_{i}^{2}\cos^{2}\theta+b_{i}^{2}\sin^{2}\theta}}}$.
The impact parameter is related to the impact angle by $\sin\theta=b/(\xi_{1}+\xi_{2})=b/d.$
The distances from the centre of mass are $r_1 = m_2d/(m_1+m_2)$ and $r_2 = m_1d/(m_1+m_2)$, and the spin rate is 
\begin{equation}
\Omega=\frac{5\mu d v_{\rm esc}\sin\theta +m_{1}(a_{1}^{2}+b_{1}^{2})\Omega_{1}+m_{1}(a_{1}^{2}+b_{1}^{2})\Omega_{2}}{5\mu d^{2}+m_{1}(a_{1}^{2}+b_{1}^{2})+m_{2}(a_{2}^{2}+b_{2}^{2})}\label{eq:p3}.
\end{equation}

Fig. \ref{fig:spin} shows the spin period dependence on the impact angle for the typical parameters of MU$_{69}$. The spin period is $P=2\pi/\Omega$, where $\Omega$ is given by Eq. (\ref{eq:p3}), with no internal rotation $\Omega_1=\Omega_2=0$. We see that an impact angle of $\sim 40^{\circ}$ gives the observed spin period. We have preformed hydrodynamical simulations based on the code of ref.\cite{SchaferEtAl-2016} that qualitatively agree with our assumptions and reproduce similar results. Typical classical KBO objects could have primordial spin periods\cite{T2014} with comparable contributions to the angular momentum budget. Recently, ref.\cite{TS2019} found that the average cold classical KBO spin periods iss $9.48\pm 1.53$ h. Generally, there is no reason for the spin vectors of each body to be correlated, so on average the contribution is zero. In extreme cases, the spin vector of both objects could be aligned or anti-aligned with the orbital angular momentum. In these cases, a wide range of impact angles and spin configurations are possible, resulting in the observed spin period after the collision. For a typical period of $10\ \rm h$, the impact angle is $\sim 20^{\circ}$ in the aligned case, and $\sim 70^{\circ}$ in the anti-aligned case.

\section*{N-body stopping conditions and tests}
We impose a stopping condition  that the distance between the two bodies is less than their mutual radius. During the non-secular highly eccentric passage, the change in the pericentre $q$ is much faster than the \textit{inner} orbital period (it is actually the definition of the non-secular regime), hence the orbital elements are not reliable at this stage. Once the simulation stops it records the orbital elements \textit{at impact} which we use for our statistics, but are not involved in the stopping condition. From the output we know the closest approach at impact. We have tested the stopping condition by varying it to be slightly smaller or larger than the $q$ we found in the first run. Indeed when the stopping condition was below $q$ the objects did not collide and continued running. We thereby concluded that the collision is physical and reliable. 

Extended Data Table 1 shows the merger fractions from the simulations. The merger fraction $f_i$ is the total number of mergers divided by the initial number of runs, multiplied by the relative fractions of the inclination distribution, assuming that no mergers occur outside the sampled inclination distribution. The fraction $f_{\rm 80-100}$ is calculated in the same way, only that the mergers considered are where the mutual inclination during the merger is within the designated boundaries of $80^{\circ}-100^{\circ}$. For example, for $\alpha=0.2$, the merger fractions is $78/200=0.39$. Multiplied by the range of the inclination distribution it is $f_i = 0.39 \times 0.3 \approx 0.12$. For $\alpha=0.3$, the merger fractions is $99/250=0.396$. Multiplied by the range of the inclination distribution it is $f_i = 0.396 \times 0.4 = 0.158$.

\begin{table}
\textbf{Extended Data Table 1| Merger rate of the binaries in the non-secular regime.}  Top to bottom: $N_m$ - total number of mergers. $N_{\rm 80-100}$ - number of merger with inclination $80^{\circ}<i<100^{\circ}$. $f_i$ - total merger fraction normalized to the inclination sampling rate. $f_{\rm 80-100}$ - the same merger fraction, only with merger at inclination $80^{\circ}<i<100^{\circ}$.
\begin{center}
\begin{tabular}{|c|c|c|c|c|}
\hline
 & $0.2$ & $0.3$ & $0.4$ & U\tabularnewline
\hline
\hline
$N_{{\rm m}}$ & $78$ & $99$ & $114$ & $101$\tabularnewline
\hline
$N_{{\rm 80-100}}$ & $9$ & $12$ & $18$ & $15$\tabularnewline
\hline
$f_{i}$ & $0.12$ & $0.158$ & $0.18$ & $0.16$\tabularnewline
\hline
$f_{80-100}$ & $0.014$ & $0.015$ & $0.029$ & $0.02$\tabularnewline
\hline
\end{tabular}
\par\end{center}
\label{tab:1}
\end{table}

\section*{Impact modelling}
We perform hydrodynamical collision simulations using our SPH code\cite{SchaferEtAl-2016}, which treats self-gravity, gas, fluid, elastic, and plastic solid bodies with material strength, including a porosity and fracture model that can be applied for small-body collisions\cite{WandelEtAl-2017,HaghighipourEtAl-2018}. In order to treat numerical rotational instabilities, a tensorial correction scheme\cite{Speith-2006} is implemented.  The \emph{miluphCUDA} code is implemented with CUDA, and runs on graphics processing units (GPU), with a substantial $\sim10^{1}-10^{2}$ improvement for single GPU compared to a single CPU. The code has already been successfully applied to several studies involving impact processes \cite{DvorakEtAl-2015,MaindlEtAl-2015,HaghighipourEtAl-2016,SchaferEtAl-2017,WandelEtAl-2017,BurgerEtAl-2018,HaghighipourEtAl-2018,MalamudEtAl-2018,mal1,mal2,mal3}. 

For the porosity treatment, we implement the \textit{P$-\alpha$} model\cite{Herrmann-1969,CarrollHolt-1972},
where the pores are much smaller than the spatial resolution and cannot be modelled explicitly. Here, the total change in the volume depends on both the compaction/collapse of the pore space, and the compression of the solid material which constitutes the matrix. The dependence is expressed in terms of the porous material pressure $P$ and density $\varrho$ as $P/\varrho = P_s / \varrho_s$, where $P_s$ and $\varrho_s$  are the pressure and density of the \textit{solid} matrix material, respectively. The distention parameter $\bar{\alpha} = \varrho_s/\varrho$ is the ratio between the solid matrix material and the porous material densities, and relates to the porosity $\psi$ via $\psi=1-1/\bar{\alpha}$. 
For the solid matrix material we use the Tillotson equation of state (EOS) parameters\cite{JutziEtAl-2009} with a reduced bulk modulus of $A=2.67\times 10^8$ Pa (leading term in the EOS) in order to consistently take into account the smaller elastic wave speeds in porous compared to solid materials, and in compatibility with ref.\cite{LeleuEtAl-2018}. Our matrix density is chosen to be $2\ \rm g\  cm^{-3}$, about the same as that of ref.\cite{JutziEtAl-2017}, which leads to a $50\%$ porosity for our fiducial bulk density for $\rm MU_{69}$, $1\ \rm g\  cm^{-3}$. The matrix density and the initial porosity are both in rough agreement with what might be expected from an object of this origin and size range. The former in particular, constraints the rock/ice mass ratio to be about 3-4 (depending on the exact choice of silicate grain density), which could be compatible with this kind of object\cite{RotundiEtAl-2015,MalamudPrialnik-2015,LorekEtAl-2016,FulleEtAl-2017}. We note however that given the uncertainties involved, we only seek to obtain a rough estimate of the density which will permit us to test our working hypothesis. We then also test some simulations with 75\% porosity and half the previous bulk density, in order to establish qualitative differences among these two setups.

For collisions between small, porous bodies, compressibility is limited by the crush curve for $\bar{\alpha}$ for typical pressures, rather than the Tillotson EOS parameters. We thereby  choose three sets of crush curve parameters, similarly to ref.\cite{JutziEtAl-2017}, using a simple quadratic crush curve\cite{JutziEtAl-2008}:
\begin{equation}
\bar{\alpha}=1+(\bar{\alpha}_0-1)\frac{(P_s-P)^2}{(P_s-P_e)^2},
\label{eq:quadritic}
\end{equation}
where $\bar{\alpha}_0=2$, $P_e$ is the transition pressure between the elastic and plastic regimes and $P_s$ is the pressure of full compaction. Both $P_e$ and $P_s$ are listed in the Extended Data Table 2. Since ref.\cite{JutziEtAl-2017} treats comet 67P/Churyumov-Gerasimenko, which belongs to a class of much smaller and active objects, we assume it is likely to be a lot more fluffier and porous. Hence our low-strength crush curve values correspond to the high-strength values from\cite{JutziEtAl-2017}, and taking the same modeling approach, we then increment the parameters in each subsequent model by one order of magnitude.

\begin{table}
\textbf{Extended Data Table 2| Crush curve, plasticity and fragmentation parameters.}
\begin{center}
\begin{tabular}{|c|c|c|c|c|}
\hline
Type 			& $P_e$ (Pa) & $P_s$ (Pa) & $Y_0$ (Pa) & $Y_T$ (Pa) \tabularnewline
\hline
Low-strength 	& $10^4$ 	& $10^6$ 	& $10^4$	 & $10^3$ \tabularnewline
\hline
Medium-strength & $10^5$ 	& $10^7$ 	& $10^5$ 	 & $10^4$ \tabularnewline
\hline
High-strength 	& $10^6$ 	& $10^8$ 	& $10^6$ 	 & $10^5$ \tabularnewline
\hline
\end{tabular}
\par\end{center}
\label{tab:2}
\end{table}

Fracture and brittle failure are treated using the Grady and Kipp fragmentation prescription\cite{GradyKipp-1993,BenzAsphaug-1994,BenzAsphaug-1999}, which is based on randomly distributed flaws in the material following a Weibull distribution with material-dependent parameters. The lowest activation threshold strain derived from the Weibull distribution, is given by $\epsilon={kV}^{-1/m}$, where $V$ is the volume of the brittle material and $k$ and $m$ are the material dependent Weibull parameters. We adopt $m=9.5$ for pressure-dependent failure\cite{Jutzi-2015} . The volume is calculated given the $\rm MU_{69}$ binary dimensions. From the material strength parameters $K$ and $G$, Young's modulus $E$ may be calculated as $E=(9KG)/(3K+G)$. Here $K=2.67 \times 10^8$ Pa, as is the leading term in the Tillotson EOS, and $G=1.6 \times 10^8$ Pa. Finally, for undamaged material $\epsilon=Y_T/E$, where $Y_T$ is the tensile strength in Table \ref{tab:2}\cite{JutziEtAl-2017}. Therefore, $k$ may be extracted and equals to $k=10^{47}, 2\times10^{39}, 2 \times 10^{28}\ \rm m^{-3}$ for the low, medium and high-strength material setups, respectively. Damage accumulates when the local tensile strain reaches the activation threshold of a flaw.

For the plasticity model we use a pressure-dependent yield strength\cite{CollinsEtAl-2004} with the implementation of ref.\cite{Jutzi-2015}. The yield stress $Y_i$ is different for damaged/intact material. For intact material the yield stress is $Y_i=Y_0+\mu_i P/\left(1+\mu_i P / (Y_M-Y_0)\right)$,
where $Y_0$ is the cohesion (see Table \ref{tab:2}), $\mu_i$ is the coefficient of friction and $Y_M$ is the shear strength at $P=\infty$. We adopt $\mu_i=1.5$\cite{Jutzi-2015} and a typical $Y_M=1.5 \times 10^9$ Pa\cite{BenzAsphaug-1999}, which is appropriate for an object composed of ice, rock and organics. For $P=0$, we recover the pressure-independent form $Y_i=Y_0$. For damaged material the yield stress is $Y_d=\mu_d P$, where $\mu_d$ is the coefficient of friction of the damaged material. Here we take $\mu_d=0.6$ following ref.\cite{Jutzi-2015}, thus fully damaged particles still undergo some shear stress.
\begin{figure}
\begin{centering}
\includegraphics[width=16cm]{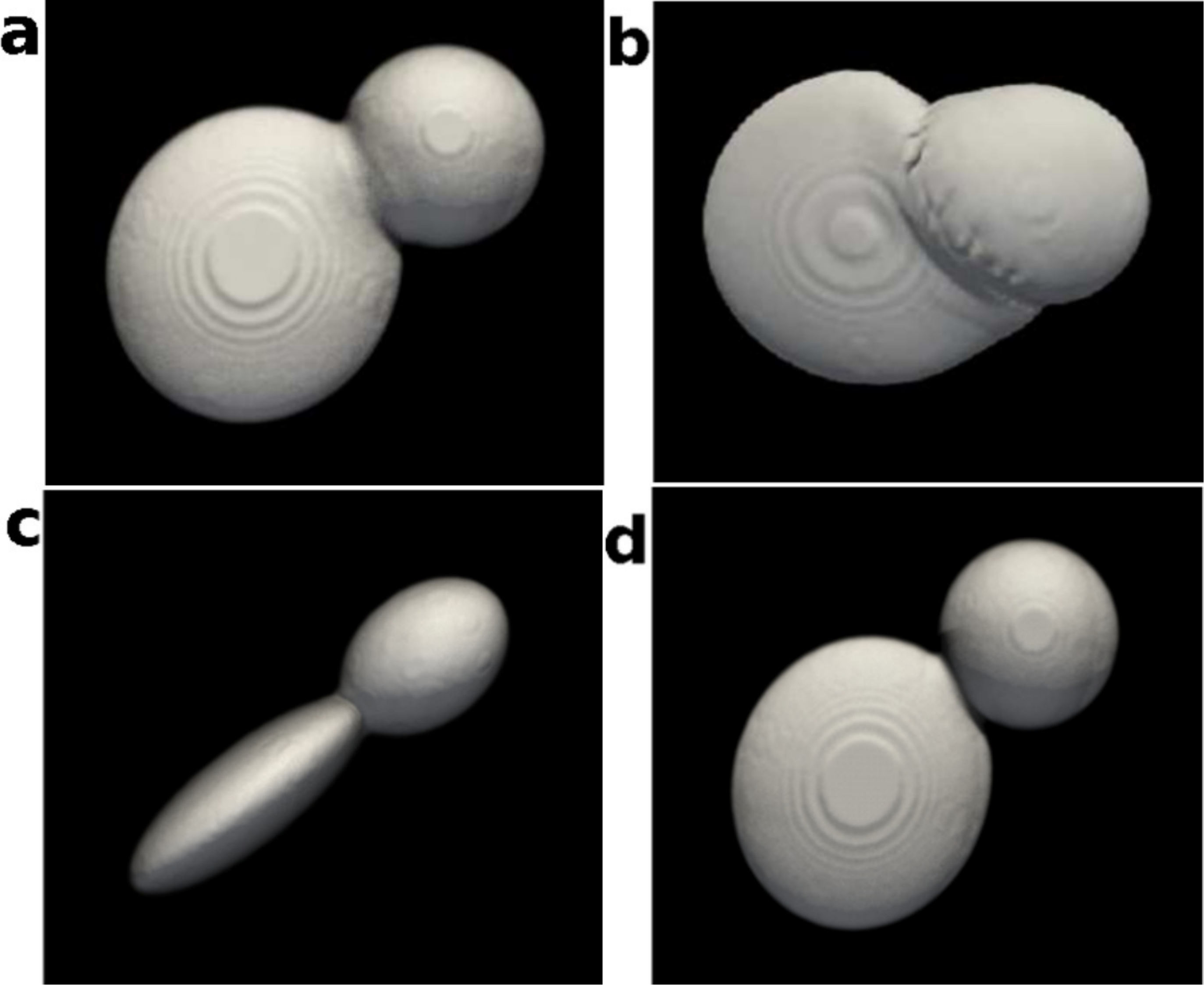}

\par\end{centering}

\label{fig:supp5} 
\caption{\textbf{Extended Data Fig. 1| Additional results of the collision models. (a),} $40^{\circ}$ impact angle, medium-strength material. \textbf{(b),} $40^{\circ}$ impact angle, low-strength material. \textbf{(c)-(d),} Low $0.5 \ \rm g \ cm^{-3}$ density model with impact angle of $55^{\circ}$ and medium-strength material. \textbf{(c),} Edge view. \textbf{(d),} Face view.}
\end{figure}

 \begin{figure}
\begin{centering}
\includegraphics[width=7cm]{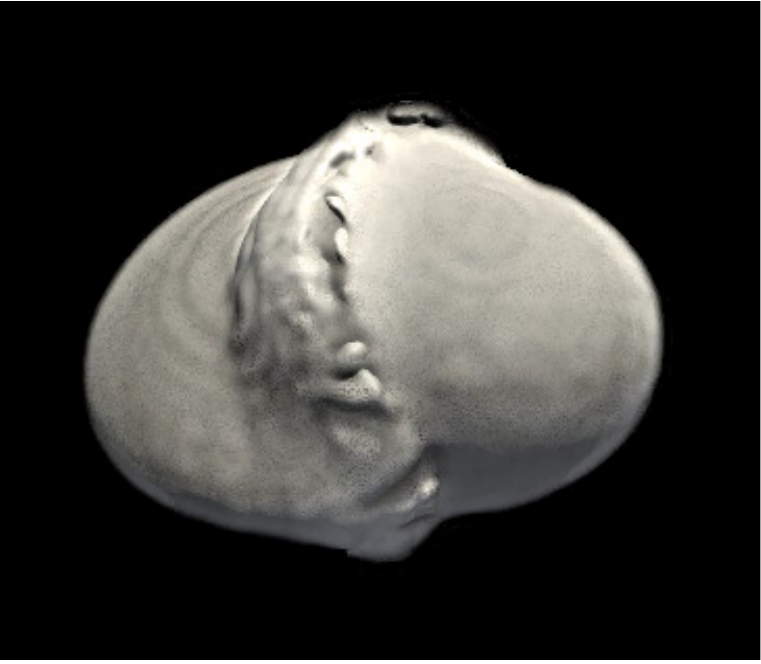}
\par\end{centering}

\label{fig:supp7}
\caption{\textbf{Extended Data Fig. 2| Additional results of the collision models.}  $5^{\circ}$ impact angle, high-strength material, large $v=10v_{\rm esc}$ escape velocity.}
\end{figure}

Extended Data Figure 1 shows additional results of our simulated impacts. We obtain the $\rm MU_{69}$ rotation period, using the nominal density of $1\ \rm g\  cm^{-3}$, only when using the high-strength material parameters. Medium \textbf{(a)} or low \textbf{(b)} strength materials deform $\rm MU_{69}$ and do not produce the observed shape of a gently merged contact binary. If the nominal density is halved ($0.5\ \rm g\  cm^{-3}$), the $v_{\rm esc}$ impact velocity is lower, which produces less deformation in our simulations. Even medium-strength material parameters generate a gently merged contact binary for virtually all impact angles \textbf{(c)-(d)}. Here, we used the $55^\circ$ impact angle, for which the observed spin period of $\rm MU_{69}$ is approximately obtained. In Extended Data Fig. 2 the shape is significantly deformed after the collision if the impact velocity is large $v = 10\ v_{\rm esc}$, using high-strength material parameters. The same velocity with weaker material parameters further leads to a complete disruption.

Our simulations were performed for a grid of impact angles, assuming pre-alignment of the two lobes. Simulating higher impact angles and low and medium strength materials causes the two lobes to interact in other ways: In some cases they hit and create contact craters, and then roll on top of each other. In other cases they stroke through each other and bounce instead of roll, and then return to re-collide following a (now) shorter orbital period. These formation channels may also generate compatible shapes but (so far) not yet the exact $\rm MU_{69}$ rotation period. In fact, a full investigation of the collision phase space must also include pre-self-rotation of each lobe, in addition to inclined hits. This entails a huge collision phase space exceeding the scope of this work, and necessitates a dedicated hydrodynamical study. Preliminary results (in preparation) indicate that such an approach may yield more channels which might generate the unique orientation of $\rm MU_{69}$, besides the successful cases (of a pre-aligned binary components) shown here.

 Our standard resolution is $5 \times 10^5$ SPH particles. We have additionally preformed simulations with $10^5$ and $2.5\times 10^5$ particles. Test Simulations were performed on the 'TAMNUN' GPU cluster, at the Technion Institute in Israel, and production runs on the bwForCluster BinAC, at T{\"u}bingen University.

\end{document}